# Model for initiation of quality factor degradation at high accelerating fields in superconducting radio-frequency cavities

A Dzyuba[1, 2], A Romanenko[1], and L D Cooley[1]

[1] SRF Materials Group, Technical Division, Fermi National Accelerator Laboratory, Batavia, IL 60510, U.S.A.

[2] Novosibirsk State University, Novosibirsk, Russia

**Abstract**

A model for the onset of the reduction in SRF cavity quality factor, the so-called *Q*-drop, at high accelerating electric fields is presented. Since magnetic fields at the cavity equator are tied to accelerating electric fields by a simple geometric factor, the onset of magnetic flux penetration determines the onset of *Q*-drop. We consider breakdown of the surface barrier at triangular grooves to predict the magnetic field of first flux penetration $H_{pen}$. Such defects were argued to be the worst case by Buzdin and Daumens, [1998 *Physica C* **294** 257], whose approach, moreover, incorporates both the geometry of the groove and local contamination via the Ginzburg-Landau parameter κ. Since previous *Q*-drop models focused on either topography or contamination alone, the proposed model allows new comparisons of one effect in relation to the other. The model predicts equivalent reduction of $H_{pen}$ when either roughness or contamination were varied alone, so smooth but dirty surfaces limit cavity performance about as much as rough but clean surfaces do. Still lower $H_{pen}$ was predicted when both effects were combined, *i.e.* contamination should exacerbate the negative effects of roughness and vice-versa. To test the model with actual data, coupons were prepared by buffered chemical polishing and electropolishing, and stylus profilometry was used to obtain distributions of angles. From these data, curves for surface resistance generated by simple flux flow as a function of magnetic field were generated by integrating over the distribution of angles for reasonable values of κ. This showed that combined effects of roughness and contamination indeed reduce the *Q*-drop onset field by ~20%, and that that contamination contributes to *Q*-drop *as much as* roughness. The latter point may be overlooked by SRF cavity research, since access to the cavity interior by spectroscopy tools is very difficult, whereas optical images have become commonplace. The model was extended to fit cavity test data, which indicated that reduction of the superconducting gap by contaminants may also play a role in *Q*-drop.

**1.Introduction.**

Superconducting radio-frequency (SRF) cavities are enabling technology for high-power linear accelerators. The state of SRF cavity art has evolved to a high degree, where highly pure niobium cavities are formed, welded, and chemically processed according to complex and highly proscriptive recipes. After receipt of a welded cavity from an industry vendor, the present widely used laboratory processing recipe includes steps of deep material removal to eliminate defects from forming and welding, ultrasonic rinsing to remove chemical residues, high-temperature baking under high vacuum to outgas hydrogen, final surface processing to produce a smooth mirror-like surface, high-pressure rinsing with ultra-pure water, drying and assembly in a clean room, and a low-temperature post-process bake under vacuum. At the end of this process, cavities are tested for the maximum accelerating electric field $E_{acc}$ to which a high quality factor $Q \sim 10^{10}$ can be maintained.

In principle, the maximum value of $E$ is related to the maximum magnetic field $H_{RF}$ that can be sustained by the surface barrier against flux penetration at the high magnetic field regions of the cavity [1]. Electric field is high near the axis of the SRF cavity, where it accelerates a particle beam, and the associated magnetic field is highest at the interior equatorial surfaces of the cavity for typical accelerating modes. Thus, increasing $E$ implies increasing $H_{RF}$ according to a simple geometric ratio given by the shape of the cavity, and it is equivalent to discuss $Q(E)$ in terms of $Q(H_{RF})$ and vice-versa. The RF limit for an ideal flat niobium sheet with a perfect surface barrier is thought to be given by the superheating field $H_{SH}$ [1], which was calculated for a few limiting cases of Ginzburg-Landau parameter κ [2,3,4], but an exact value for niobium at 2 K has not been analytically calculated. One of the theoretical estimates gives $H_{SH} \approx 1.2 H_c$ for κ ~ 1. For very pure niobium, $\mu_0 H_c \approx 180$ mT at 2 K, and SRF cavities have approached 200 mT surface field [5,6], demonstrating the superheating nature of the phase boundary. For cavities with the so-called TESLA shape, upon which much of the recent literature is based, the ratio of $\mu_0 H_{RF}$ to $E$ is about 4.37 mT per (MV m$^{-1}$), implying a maximum theoretical electric field of ~45 MV m$^{-1}$.

Since most RF tests achieve $E_{acc}$ > 20 MV m$^{-1}$, $H_{RF}$ is easily 50% or more of $H_c$. Surface currents then approach the depairing limit ~$H_c/\lambda$ there, where λ is the magnetic penetration depth. This implies that once the surface barrier is overcome, dissipation due to violent vortex motion immediately ensues, since the strongest flux-pinning forces typically sustain in critical currents not more than 10% of $H_c/\lambda$. The vortex motion produces dissipation and a pronounced drop of $Q$ followed by local heating [7]. Thus, a signature of cavity tests is a phenomenon called "$Q$-drop", where $Q$ declines steeply above a clearly discernable onset field $E_{onset}$ at which the surface barrier breaks down. When the forming, welding, and processing recipe above is executed to a high degree of perfection, $E_{onset}$ can exceed 35 or even 40 MV m$^{-1}$, and it is difficult to detect a particular location where breakdown occurs. The range of performance should then be related to the general quality of the surface barrier throughout the cavity, and this should be traceable back to the processing data. Optimization of cavity performance should then be straightforward. Sometimes, however, significant flaws occur, which produce $Q$-drop at much lower electric fields, with local heating being detected at locations correlated with the location of the flaw [8].

This article explores the mechanism of $Q$-drop above 25 MV m$^{-1}$, although it may also be applicable to $Q$-drop at lower values of $E_{onset}$ if the mechanism by which a certain flaw initiates $Q$-drop is through severe impairment of the surface barrier. In a practical sense, a model that combines effects of both topography and contamination on the breakdown of the surface barrier would be very helpful. The obvious topographical features in SRF cavities are grain boundaries, which have a step-like or groove-like profile based on various imaging techniques [9,10]. Occasionally large flaws, such as pits and deep scratches, are observed, but because they have sizes on the order of $10^3$ to $10^4 \lambda$, sub-features are thought to provide the initiation of $Q$-drop [11]. Curiously, grain boundaries are often observed inside these large defects. Contaminants can penetrate into boundaries [12,13], and the action of contaminants could transform otherwise benign boundaries into obstructions of shielding current and pathways for vortex penetration. Either of these conditions would modify the surface barrier. Interstitial contaminants beneath a smooth niobium surface tend to increase electron scattering [1], which directly increases κ, and they also bind dislocations and make recovery anneals less effective at restoring high conductivity. Breakdown of the surface barrier by sub-surface interstitial contaminants has been well documented for investigations of very pure V and well-annealed Nb-Ta with slight exposure to oxygen [14,15].

Clearly, therefore, both topography and contamination influence cavity behavior, but their relative importance is difficult to assess. The relative importance of topography vs. contamination is the scientific objective of this article. Such an assessment would be valuable to validate and optimize the different processing steps. Access to the cavity interior to gain spectroscopic data is extremely difficult, so destruction of the cavity (which represents a significant investment) is required to study contamination. Preparation of coupon samples is difficult to make relevant to actual cavity processing conditions since the thermal gradients, flow patterns, and mechanical treatments such as deep drawing are difficult to simulate exactly. While topographical data can be obtained directly from cavities by replicas [16,17], and inferred from images obtained by new optical inspection systems [9], so far the primary use of cavity inspections has been to debug gross processing defects [8,18], and discussion of fine-scale information is only just beginning.

We structure this article as follows: In section 2, we review the models of Aladyshkin *et al.* and Buzdin and Daumens [19,20], which assess not only topographical effects but incorporate contamination via the Ginzburg-Landau parameter $\kappa$. In section 3, we analyze profilometry data for polycrystalline and single-crystal niobium prepared by both electropolishing (EP) and buffered chemical polishing (BCP), including one sample directly extracted from a SRF cavity for which $Q(E)$ is known. The sharpest features, represented by the tails of an angular distribution, are those which initiate flux penetration and the onset of $Q$-drop. We use this analysis in section 4 to fit $Q(E)$ data for a cavity, which suggests that surface contamination is much stronger than expected. In section 5, we discuss other aspects of our model, including implications of our results for the present methods used to polish the RF surface of cavities. In section 6 we present conclusions.

## 2. Reduced surface barrier at grooves

The ideal surface can be described by the Bean-Livingston model [22], where the onset of flux penetration with increasing field is delayed by the presence of a surface barrier, despite the fact that the net energy difference for flux penetration is negative for field higher than the lower critical field $H_{c1}$. The presence of the surface barrier is due to the interplay between two electromagnetic forces acting on a single vortex; attraction of the vortex to the surface (image force), and repulsion due to screening currents pushing the vortex toward the superconductor interior. For ideally flat surfaces, the penetration field was found to be close to $H_c$.

There has been substantial work on the impact of topographical features on the surface barrier. The strongest impact on the surface barrier occurs when the field runs in the direction of the feature and Meissner currents are diverted around the obstruction, creating regions of current concentration near the defect boundary. The concentration of current at the groove apex enhances the force tending to drive a vortex into the superconductor interior, and the deflection of the surface from parallel reduces the image force seen by the vortex; this is why the surface barrier is reduced. Here, we consider that 2-dimensional features, which uniformly concentrate shielding current toward the interior of the superconductor, act more strongly than 1-dimensional holes, which allow current to redistribute along the surface as well as beneath the defect. Aladyshkin *et al.* [19] and Buzdin and Daumens [20] explored the effects of a wedge-shaped groove with field running in the direction of the groove, so that Meissner currents concentrate near the crack tip. They assessed the field of first flux penetration $H_{pen} = \gamma H_c$, and found that the surface-barrier reduction factor $\gamma$ falls with increasing sharpness of the groove. Here, the groove depth is larger than $\lambda$ so that it impacts the Meissner current. The maximum suppression of the surface barrier was found for the limit of crack being a half-planar boundary (zero width at the surface), where $\gamma = 2\,\beta^{-1}\,\kappa^{-1/2}$ with $\kappa = \lambda/\xi$ being the Ginzburg-Landau parameter, $\xi$ is the superconducting coherence length, and $\beta$ being a factor of order 1. This situation is analogous to a contaminated grain boundary or gap such that very little current can cross the boundary. Vodolazov [21] explored rectangular troughs in a similar geometry as the triangular groove of Aladyshkin *et al.* and Buzdin and Daumens. Vodolazov used numerical methods to obtain $\gamma \approx \kappa^{-1/3}$ for rectangles with width $\xi$, with $\gamma$ increasing with width due to the disturbances at the corners spreading apart. For either model, since $\kappa$ is 2 to 7 for SRF niobium, reduction of the surface barrier by a factor of ~2 is likely, although heavier surface contamination could decrease $\gamma$ substantially by driving up $\kappa$ locally.

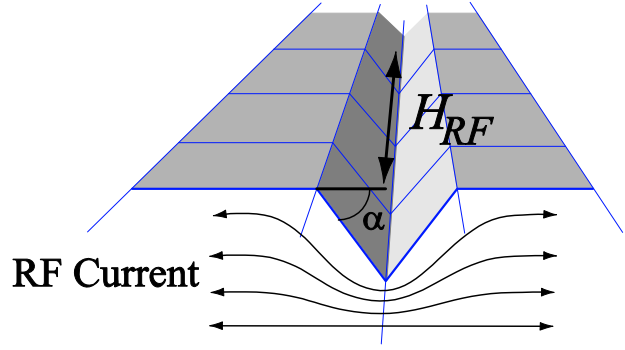

Figure 1: Geometry of the problem and definition of the angle α for the triangular groove. The defect produces a concentration of shielding current beneath the deepest point.

We summarize the approach of Buzdin and Daumens [20] in this section. It is valid in the limit that the Laplace equation can be solved, i.e. for groove widths comparable to λ. Their solution used a conformal mapping technique based on the transformation of shapes, which preserves angular information. We change the definition of angle α to accommodate later topography measurements by the profilometer, where a perfectly sharp groove with 0 angle between the walls in [20] corresponds to π/2 in our configuration. Figure 1 shows the geometry of the groove. The first penetration field is

$$H_{pen} = \left(\frac{\xi}{\lambda}\right)^{1-\frac{\pi}{\pi+2\alpha}} H_c , \quad (1)$$

implying that

$$\gamma = \kappa^{\frac{-\alpha}{\pi/2+\alpha}} , \quad (2)$$

Eq. (2) has the limit γ = κ$^{-1/2}$ for α → π/2, as stated earlier. Figure 2 shows a calculation of γ plotted as a function of angle for κ = 7 (red line) and 2 (blue line) to demonstrate the relative differences for the range of purity expected for processed niobium cavities.

### 3. Niobium surface profiles

*3.1 Profilometry of niobium samples prepared by cavity polishing techniques*

Since the first flux entry field at each wedge-like surface step given by Eq. (1) depends on the angle α, it is instructive to obtain a distribution of angles on real niobium samples to ascertain those features which admit flux first and initiate *Q*-drop. In this section, we compare surface profiles for niobium subjected to different surface chemical treatments to infer penetration fields encountered in niobium cavities.

Two niobium samples were analyzed with a stylus profilometer: a fine-grained (FG) sample (~50 μm grain size), and a single-crystal (SC) sample. The FG sample was typical of the starting sheets used to make SRF cavity half cells, being in a recrystallized state following an anneal in a vacuum furnace by the niobium supplier. Since cavities undergo a high-temperature bake, discussed in more detail later, effects of forming are largely removed prior to RF measurements. Thus, the FG sample is a reasonable approximation of the recovered and recrystallized SRF cavity material.

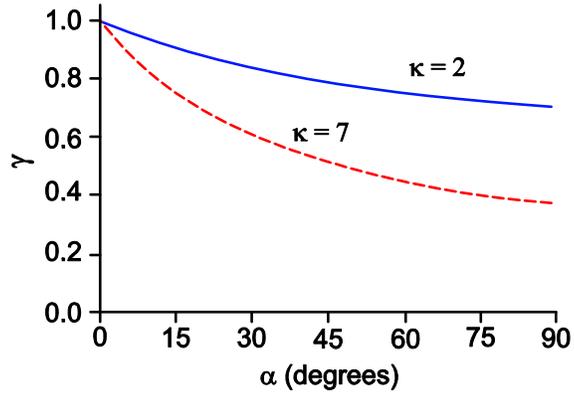

Figure 2: Calculation of penetration fields as a function of increasing angle $\alpha$ for $\kappa = 7$ (red dashed curve) and 2 (blue solid curve).

BCP was applied to remove about 100 μm of material for each sample. A profile was then acquired for this state for each sample. The FG sample profile is shown in figure 3. Next, electropolishing was used to remove 115 μm of additional material from both samples, and a second profile was acquired for each sample. The FG EP profile is also shown in figure 3. The amount of material removed was determined by weighing the samples, and material was removed from both sides. The chemical solutions and polishing procedures were typical of those used in SRF cavity preparation. The BCP solution was a 1:1:2 mixture by volume of 48% hydrofluoric, 98% sulfuric, and 99% phosphoric acids. Acids were pre-chilled and mixed when fresh. Etching took place at 10 to 12 °C. The EP electrolyte was a solution of 1:9 by volume 48% hydrofluoric and 98% sulfuric acids. Electropolishing was carried out at 14.5 V with a current density of 500 A m$^{-2}$ and with circulation of electrolyte as low as possible to maintain temperature of ~30 °C.

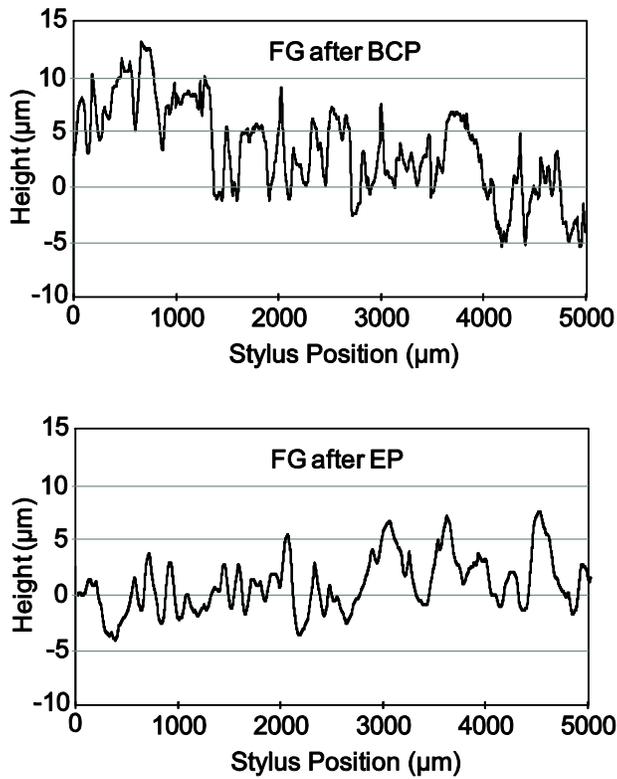

Figure 3: Profiles acquired for fine-grain (FG) niobium samples processed by BCP (top plot) and EP (bottom plot).

Profiles were acquired over 5 millimeters length using several scans to acquire a suitable statistical sample. Data shown in figure 3 was obtained for 0.5 µm step with manual leveling. The profiles show some interesting features. First, the FG profiles show significantly greater vertical range than the SC profiles (not shown). This is due to differences in the rate at which acid attacks various crystal planes for BCP, and also due to fact that the EP procedure was halted before significant micro-leveling of the FG samples could take place. The FG BCP profile displays relatively flat regions and sharp transitions between these regions. This corresponds to interiors of grains and grain boundaries. No such features are visible in the FG EP profile.

Surface angles were then calculated by moving a point-to-point linear fit through the data and recording the deviation angle from horizontal. The distribution of angles was then obtained; these data are plotted for the fine-grain specimens in figure 4. The single-crystal specimens had very narrow distributions by comparison (not shown). The broader distribution for FG BCP implies a rougher surface than for FG EP. This observation is consistent with previous observations drawn from profilometry and AFM studies [references], which show that generally electropolishing produces lower peak-to-peak and RMS surface roughness than BCP. It should be noted that in both BCP and EP cases, the steepest angles encountered are close in value, although the relative frequency is different. This point will be expanded in the next section.

In all cases a reasonable fit to the distribution of angles $n(\alpha)$ was provided by the Lorentzian

$$n(\alpha) = C \frac{w}{w^2 + \alpha^2} \quad (4)$$

where $w$ is the width of the distribution at half-maximum and $C$ is a constant. In figure 4, the parameter $w$ for FG BCP is 6.3° and for FG EP it is 2.3°. The Lorentzian distribution of angles is different from the Gaussian distribution used in a previous model [23] to describe $Q$ vs. $E$ curves. Indeed, in some cases a Gaussian distribution also gave a good fit to our data when the distribution was very narrow. The Lorentzian fit was a better fit in the tail regions for broad distributions where the largest angles occur. This region is important, because it is representative of the strongest perturbations on the surface barrier.

*3.2 Fractal nature of polished niobium surfaces*

It is important to recognize that roughness information obtained over length scales comparable to λ (~50 nm at 2 K for Nb) is beyond the limit of stylus profilometry and requires extensive analysis by atomic-force microscopy or laser scanning confocal microscopy. One way to address this difficulty is by recognizing that roughness is self-similar down to these length scales. The surface texture of many metals prepared by etching or polishing can be characterized by fractals. This is a well-known basis for understanding wear and friction, lifetimes of medical implants, and so on.

Niobium processed by EP and BCP has roughness that obeys fractal characteristics. Recent analyses [24,25] have shown that the power spectral density (PSD) of SRF niobium surface roughness scales with step frequency by a simple power law over 5 decades of inverse scan length (1/$L$). The fractal dimension is close to 2, i.e. the PSD (in nm$^2$) scales as $L^2$. Good overlap was shown between frequency ranges for both stylus profilometry and atomic-force microscopy data, where the latter extend down below 100 nm scan step. In our case, different scan frequencies were applied to the fine-grained BCP sample, ranging from 20 to 0.02 μm$^{-1}$, and the histograms of angles were analyzed in each case. Weak logarithmic increase of $w$ with step size was seen, suggesting that the distribution widths are somewhat wider when sampled with step sizes approaching λ. This is a topic for further research.

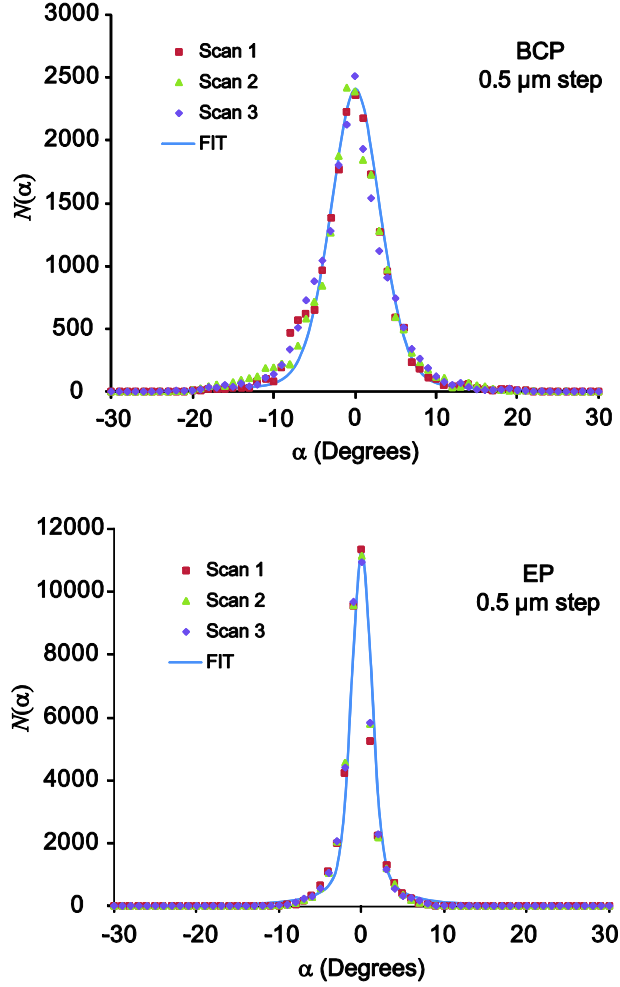

Figure 4: Histograms of the surface angles obtained from profiles on fine-grained niobium after BCP (left plot) and EP (right plot) treatments.

### 4. Model for $Q$-drop at high fields

The angular distribution of surface features can be combined with the equations for the penetration field to assess the surface resistance $R_S$ as a function of $H_{RF}$. Since $R_S$ is inversely related to $Q$,

$$R_s = \frac{G}{Q}, \quad (7)$$

the calculations of surface resistance can be used to predict the onset of $Q$-drop, where $G$ is a geometry factor (273Ω for the so-calledTesla cavity shape). We assume that flux will penetrate first at sites where the surface barrier is reduced, which then causes an increase in surface resistance due to simple oscillatory fluxoid motion driven by the Lorentz force of the RF screening current, as considered in [26].

We then consider a worst case in which the entire surface profile is composed of triangular grooves to estimate the surface resistance.

To arrive at a value for $R_s$ at particular values of $H_{RF}$ and $\kappa$, we integrate over a distribution of angles and evaluating the dissipation at regions for which $H_{pen} < H_{RF}$. At first, the most severe troughs and the regions with strongest contamination become penetrated. Increasing the RF field leads to more and more sites being penetrated and hence an increase in the fraction of the distribution contributing to the overall surface resistance. The angular integrand is expressed as a Lorentzian distribution of angles, as considered in the previous section,

$$n(\alpha)d\alpha = \frac{1}{\pi}\left(\frac{w}{w^2 + \alpha^2}\right)d\alpha. \quad (5)$$

The penetration field at a particular angle is the result of the conformal transformation for flux penetration into a triangular groove with angle α, which can be represented by re-arranging Eq (1):

$$H_{pen} = \kappa^{\frac{-\alpha}{\pi/2+a}} H_c. \quad (6)$$

For those regions which are penetrated by flux, we use the result of Rabinowitz for surface resistance due to the oscillations of a vortex that has penetrated the barrier [26]

$$\begin{aligned} R_s(\alpha, H_{RF}) &= \frac{H_{RF}}{H_c}\frac{\rho}{2\lambda} \quad \text{for } H_{RF} > H_{pen}(\alpha); \\ R_s(\alpha, H_{RF}) &= 0 \quad \text{for } H_{RF} < H_{pen}(\alpha) \end{aligned}, \quad (8)$$

where ρ is a dissipation of $10^{-14}$ Ω-m determined by thermal stability conditions [26]. We then integrate contributions to surface resistance over the angle distribution:

$$R_s^{total}(H_{RF}) = C \cdot \int_0^{\pi/2} R_s(\alpha, H_{RF}) \cdot n(\alpha) d\alpha. \quad (9)$$

The normalization constant $C$ in Eq. (9) is chosen so that $Q \sim 10^8$ when the full distribution is integrated.

Figure 5 shows the results for surface resistance increase using different distribution widths for $\kappa = 2$ and 10. By contrast, figure 6 shows calculations for $\kappa = 2, 4, 7,$ and 10 for a distribution width of 4.5°. It should be noted that typical SRF cavities exhibit residual surface resistance of 5 to 10 nΩ in steady state, not 0, before the onset of Q-drop. This baseline is not reflected in these plots. Using the normalization described above, a criterion of $Q = 10^9$ is depicted by the dot-dash line, which is a point where RF tests might be halted due to excessive dissipation.

The variation with different values for the distribution width is not large when κ is low, with a spread of ~20% at the cut-off criterion (solid curves in figure 5). Larger values of κ enhance the effect of the angle distribution (dashed curves in figure 5). The reduction of $E_{onset}$ also appears to attenuate rather quickly with broadening of the distribution above 6°, whereas gains in performance could be realized for achieving extremely narrow distributions <3°. Similarly, a rather strong variation in the onset of surface resistance is observed in figure 6 with increasing κ even for a distribution characteristic of a reasonable polish. The data show a high sensitivity of the onset of Q-drop to κ when κ < 4. We note that, while κ for SRF cavity grade Nb lies between 2 and 4 (a very good fit of the superheating field was recently obtained for κ = 3.5 [6]), measurements of κ are inferred from currents and fields passing through the bulk, which neglects the active ~50 nm thick layer for RF superconductivity at the surface. Since interstitial atom concentration is likely to be higher near the surface, κ values as high as 10 may be reasonable for the active RF layer.

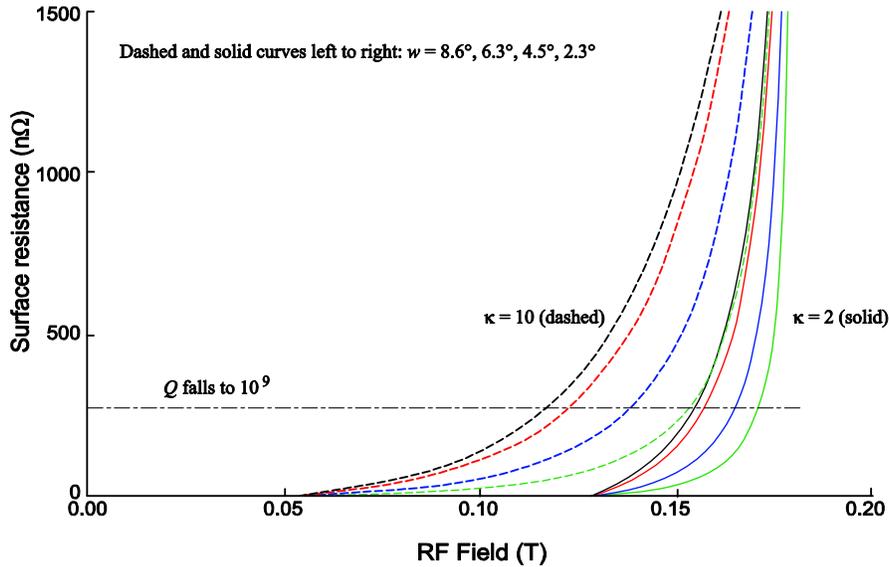

Figure 5: Calculated surface resistance as a function of RF field for κ = 2 (solid curves) and 10 (dashed curves). The widths of the distributions are decreasing from left to right for each family of curves, as indicated on the plot. For all data, $\mu_0 H_c = 0.18$ T. The dot-dash line suggests the point at which $Q$ falls to $10^9$, a criterion where SRF cavity tests are often terminated.

## 6. Discussion

*6.1. Which is more important, roughness or contamination?*

The results of the model suggest that contamination plays at least as significant a role in determining the $Q(E)$ behavior at high field as does roughness. The span of solid curves representing low contamination in figure 5 is quite small, with a ~10% variation of the value of $E_{onset}$. This is much less than what is presently observed [27-29], where cavities that are processed by a consistent set of EP parameters and have no obvious flaws exhibit a range of $E_{onset}$ that is ~30% of the mean, from ~30 to ~40 MV/m at 2 K. When contamination is heavy, given by the dashed curves in figure 5, the full range of angular distributions explored does simulate the wide range of performance seen in practice. The differences are most pronounced in the lower portions of the curves, where the contributions of the tails of the angular distributions are strongest.

$E_{onset}$ was quite sensitive to κ, especially for small $w$. For the data shown in figure 6, a cavity with $w$ = 4.5° and $E_{onset}$ = 35 MV m$^{-1}$ (or $\mu_0 H_{RF} \approx 0.15$ T) must be very clean, with κ = 2. Moderate contamination, sufficient to drive κ up to ~7, produces a low-level rise in $R_s$ at fields as low as 0.10 T, implying a degradation of $E_{onset}$ by ~30%. Since the RF activity is ~50 nm deep and is not characterized by bulk measurements of the niobium superconductivity, values of κ higher than those deduced from bulk measurements could be more appropriate to this top layer, which is why we chose to simulate κ values as high as 10. Indeed, the fact that thermodynamics permits interstitial atom concentrations a few percent suggests that exposure of a fresh Nb surface to air or water should result in substantial contamination. Surface studies [1,13,27,30–33] support aspects of this "pollution" model. As discussed further below, a change in surface contamination might be the key mechanism beneath the improvement of cavities for low-temperature (120 °C) baking in vacuum.

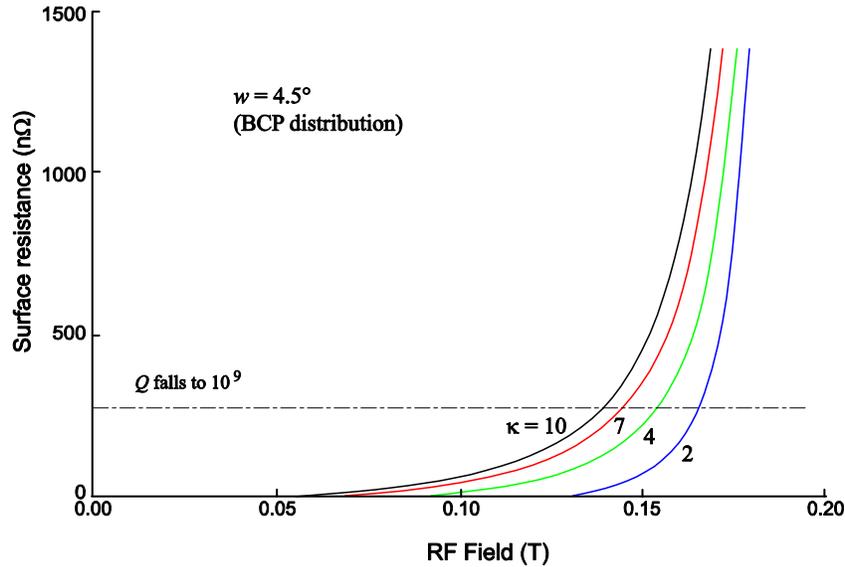

Figure 6: Surface resistance as a function of RF field for $\kappa$ = 2, 4, 7, and 10 (from right to left). The width of the distribution is 4.5°, characteristic of a BCP sample, and $\mu_0 H_c$ = 0.18 T. The dot-dash line indicates where $Q$ falls to $10^9$.

Combined increases of both $\kappa$ and angular distribution breadth exacerbate the individual trends for changing either variable alone. Evidently, both substantial contamination as well as some breadth in the angular distribution is required for the model to predict the range of $E_{onset}$ actually seen in practice. This is explicit in figure 5. An obvious conclusion is that *synergistic* improvement of *both* variables is desirable. Stating this differently, it might be concluded that *both* roughness and contamination are active to comparable degrees at limiting SRF cavity performance. A difficult question to answer is how to provide metrics that can be used to gauge improvement. As mentioned in the Introduction, topographical information for cavities is now relatively easy to obtain via the use of optical inspection systems and silicone replicas [17,18]. Proposed systems could add interferometry, stereo imaging, gloss measurements, and computational techniques to acquire roughness data *in-situ* and avoid the replica process altogether. Quantitative information about contaminants is more daunting. Spectroscopy requires the dissection of cavities to provide samples to experimentalists.

Further guidance from coupon experiments may have limited value. Coupon simulations are useful for understanding the differences between BCP and EP, and good correlation between cavity and coupon surfaces prepared by BCP has been noted [34]. On the other hand, the forming and welding conditions, rotation of cavities during EP processing, temperature gradients, and mixing action are all difficult to simulate at a coupon level. It is also quite difficult to conduct measurements of $Q(E)$ on coupons, although surface resistance can be probed by other RF techniques. Comparison of results may also be difficult: Coupon experiments conducted at Fermilab suggest that the roughness is somewhat higher in cavities than for coupons, which suggests that the value of $w$ assessed in figure 4 might be too low to represent real cavity surfaces.

*6.2. Why does electropolishing with low-temperature baking result in better performance?*

The ability of the model proposed here to compare the effects of roughness with those of contamination the can be directly applied to improving cavity processing. Electropolishing has become preferred over buffered chemical polishing for preparing the RF surface [1,31] because of the apparent correlation between higher $E_{onset}$ with lower surface roughness, as measured by stylus profilometry and atomic force microscopy [24]. This correlation breaks down, however, when EP is not followed by a final vacuum baking at ~120 °C, as explained further below. It would be quite useful to understand and optimize the features of the EP process that are effective for increasing $E_{onset}$, such as low roughness, while eliminating other features that are detrimental. In addition, there are curious examples where RF magnetic fields well above 150 mT were sustained despite the presence of obvious topographical flaws when low-temperature baking was applied [17,35]. Whatever the benefits of final vacuum baking are, they appear, therefore, to operate synergistically with the topography of the cavity surface. Since alternate techniques, such as tumbling or centrifugal barrel polishing, might achieve the same or even better roughness as electrochemistry can, it may be possible to avoid side-effects of electropolishing to attain cavities not limited by $Q$-drop.

Within the framework of the model, the following is a plausible explanation why EP followed by vacuum baking improves cavity performance. A typical BCP process should result in a fairly broad angular distribution, $w \sim 5°$ from figure 4, and some contamination, $\kappa \sim 5$. The model predicts a rather modest onset of $Q$-drop above about 100 mT field, e.g. by interpolating curves in figure 6, and indeed this is what is generally seen in practice—cavities processed only by BCP attain not more than 25 MV/m before they are limited by $Q$-drop. This limit is not removed by final baking; evidently the high roughness masks any changes in contamination. Electropolishing alone improves the angular distribution, with $w \sim 2.5°$ in figure 4, which pushes $E_{onset}$ up to ~30 MV m$^{-1}$ (the family of curves for $\kappa = 5$ might be interpolated between those shown in figure 5). Electropolishing combined with a high-temperature bake, followed by a final EP and a final vacuum bake at 120 °C improves both factors, $w \sim 2.5°$ and $\kappa \sim 3$, resulting in behavior similar to the rightmost solid curve in figure 5. In practice, this combination generally results in cavities with no $Q$-drop observed up to the highest fields of > 40 MV m$^{-1}$ where performance is instead limited by a quench [1].

Why should the low-temperature vacuum bake affect the Ginzburg-Landau parameter? Cavity-grade niobium contains less than a few ppm of heavy elements, except for tantalum. Surface resistance is then affected by structural defects in the metal and contamination by interstitial elements. Although all interstitial contaminants are kept to a low level by vacuum arc melting the initial niobium ingots, oxygen and hydrogen are inevitably introduced during chemical processing and baking. Oxidation is a fundamental part of both EP and BCP chemistry, and oxides naturally form when the metal surface is exposed to air or rinse water. Residual oxides may interfere with superconductivity, as suggested by recent signatures of magnetic scattering found by point-contact tunneling [36], where magnetic moments arise due to defects in the oxide layer. Oxides that penetrate into grain boundaries or dislocation clusters could also be problematic, and it may be important to provide mobile vacancies to allow healing of these defects. Perhaps this occurs during the mild final vacuum bake; however, there is no reason to suspect that EP produces a different oxide than BCP, and it is difficult to imagine why one processing route would respond to low-temperature baking while the other would not.

Hydrogen presents a different story, due to its tendency to bind with vacancies. Hydrogen moves very quickly through niobium at room temperature. Since EP takes place at 25 to 40 °C, warmer than the 10 to 15 °C range for BCP, and EP produces copious amounts of hydrogen gas at the cathode, significant amounts of hydrogen can dissolve into the niobium metal during EP but not during BCP. In terms of our model, a surface prepared by EP may thus be smoother but dirtier, whereas the surface prepared by BCP may be cleaner but rougher. Comparable RF performance prior to the final bake is, then, plausible; similar curves in figures 5 and 6 resulted for $w = 4.5°$ and $\kappa = 4$ (BCP) as for $w = 2.3°$ and $\kappa = 10$ (EP), with $E_{onset}$ values < 30 MV m$^{-1}$. Although hydrogen contamination is remediated by baking the niobium cavity at 600 to 800 °C in ultra-high vacuum for several hours, complete removal of hydrogen from the EP cavity is difficult to achieve [37], and some re-contamination is likely anyway due the use of EP for the final surface processing step.  Fortunately, dissociation of hydrogen from vacancies occurs at temperatures above ~120°C, the same range used for the low-temperature bake [38].  Release of vacancies by baking has been cited as the mechanism for an observed reduction of structural defects in samples taken from cavities processed by EP [29].  In terms of our model, the removal of defects could result in a reduction in κ during the low-temperature bake.  Since $E_{onset}$ was quite sensitive to κ, especially for low values of $w$, a significant increase of $E_{onset}$ would result.

*6.3. Fit to cavity data*

We applied Eq. (9) to fit experimental data for $R_s$ vs. $H_{RF}$ for a cavity processed by the same parameters used to generate the data in figure 4, namely BCP (100 µm Nb removed) followed by EP (an additional 100 µm removed) [30].  This cavity did not exhibit a localized breakdown due to an isolated flaw.  The fit was forced by assuming the width of the FG-EP distribution shown in figure 4, $w = 2.3°$, also applies to the cavity.  The Ginzburg-Landau parameter was allowed to vary between 2 and 10, with the best fit for κ = 5.  It was not possible to produce a good fit without reducing the value of $H_c$; in figure 7 it is 0.13 T.  Ciovati, Kneisel, and Gurevich [39] noted that the onset of $Q$-drop for a wide range of cavity treatments could be described by a local reduction of the superconducting gap near the surface by as much as 15%.  The reduction was thought to be due to dissolved oxygen (which reduced $T_c$ by ~1 K per percent oxygen [40]) and dissolved hydrogen (which affects both $T_c$ and surface resistance [41]).  The somewhat larger reduction of $H_c$ could reflect the fact that this cavity has not received baking treatments commonly used to remove surface contaminants, as discussed earlier.

It should also be noted that the model assumes that parameters are uniform throughout the entire equator region of the cavity, and that a full integration of the angular distribution is required to give the observed maximum of surface resistance.  In a true cavity, contamination is likely to be collected in pockets.  This condition might be modeled better by a range of κ values (and a range of $H_c$ values).  Also, the cavity test is usually terminated before a large value of surface resistance is reached, suggesting that only a portion of the angular distribution is active.  Improvement of the model is difficult here, since smooth but contaminated cavities (low $w$, high κ) provide losses similar to clean but rough cavities (high $w$, low κ).  Furthermore, since optical images from the cavity interior reveal grain boundaries, whereas coupon electropolishing typically produces grain-boundary free surfaces, somewhat higher values of $w$ might be more reasonable.

Our model does not take into account nonlinear effects on the surface resistance [42,43], which can significantly amplify the surface resistance pair-breaking effects of hot spots once vortex motion is initiated.  Gurevich [42] suggests that an additional factor scaling as $(H_{RF} / H_c)^2$ should be included in the data, due to the pair-breaking effects of RF currents approaching the depairing current density.

*6.4 Field enhancement effects*

This model differs substantially from previous models based on the surface roughness, for which the field is aligned transverse to surface steps and not along them.  In that case, Meissner currents flow along the defect and are not obstructed.  However, a local enhancement of the field is then thought to occur, exacerbating the effects of flux penetration through a weakened surface barrier.  Because the true topography of the cavity interior is a complex mixture of troughs and ridges, both field enhancement and surface-barrier reduction are combined.  The effects of roughness were not considered in fitting cavity data in figure 7, but they should nonetheless be present.

Knobloch et al. noted that abrupt changes in topography, such as those at step edges, terraces, grain boundaries, and peaks and valleys, produce a local enhancement of the RF magnetic field [23]. Here, the field is aligned transverse to the features, with the Meissner currents running along the features and in principle not suffering any obstruction. The enhancement factor $\beta > 1$ can be related to the angle describing the sharpness of the feature or, alternately, to the radius of curvature near the sharpest edge. This results in an effective field $H_{eff} = \beta\, H_{RF}$. This principle has been applied successfully to understand Q-drop and quench at the impact of large surface features observed in SRF cavities [11], such as abrupt edges around pits or protruding grain facets.

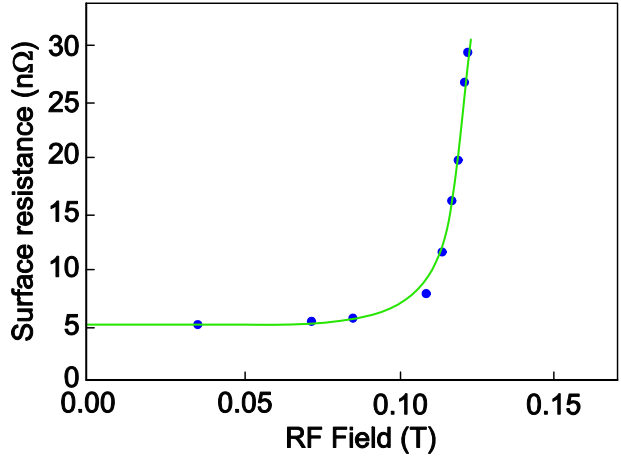

Figure 7: Fit to the surface resistance derived from cavity $Q(E)$ data. The fit was obtained for $w = 2.3°$, $\mu_0 H_c = 0.13$ T, and $\kappa = 5$.

Modification of the field enhancement model by including surface barrier and the onset of flux entry should be useful. While the field enhancement model sets the scale for the local magnitude of effective field applied to the surface barrier, the true flux penetration event still occurs over length scales comparable to $\lambda$ or about 40 nm in very pure Nb. This means that the effects of field enhancement over longer length scales must be broken down into these elementary locations where the surface barrier is defeated. Instead of $H_{pen} = \gamma\, H_c$ with $\gamma < 1$ and scaling as $\kappa^{-1/2}$ for the strongest reductions to the surface barrier, the enhancement factor would require $\beta\, H_{pen} = \gamma\, H_c$, or that the combined effects of field enhancement, roughness, and contamination result in a factor $\eta = \gamma\,/\,\beta$ that scales the maximum equatorial field. Edges of pits become rounded after electropolishing, producing $\beta \sim 1.2$ to $1.5$, comparable to the values of $(1/\gamma)$ calculated by our model. The combined factor $\eta$ can be 2 to 3. It is not surprising, therefore, that pits and bumps seen in cavities are correlated with localized quench often at 50% of the theoretical maximum field.

## 7. Conclusions

A model for the onset of the reduction in SRF cavity quality factor, the so-called $Q$-drop, at high accelerating fields was presented. The model examined the weakening of the surface barrier near triangular grooves, which are the defect types thought to most strongly affect the Meissner currents. This analysis permitted the effects of surface topography to be combined with contamination by incorporating both an angular deviation from parallel and the Ginzburg-Landau parameter. An outcome of the analysis was a rather significant reduction in the field of flux penetration when either roughness or contamination were varied alone; that is, smooth but dirty surfaces provided the same onset of flux penetration as did rough but clean surfaces. Strongest reduction of penetration field was seen when both effects were combined. Thus, contamination exacerbates the negative effects of roughness and vice-versa.

To obtain distributions of surface angles, coupons were prepared by BCP and then EP, and profiles were taken after each process. A simple model for surface resistance generated by flux flow was then proposed, and the onset of surface resistance, which is equivalent to the onset of *Q*-drop, was calculated for various values of κ and for reasonable values of the distribution width. This exercise indicated that contamination contributes to *Q*-drop *as much as* roughness. This point may be overlooked by SRF cavity research, since access to the cavity interior by spectroscopy tools is very difficult, whereas optical images have become commonplace. The model was extended to fit cavity test data, which indicated that reduction of the superconducting gap by contaminants must also be included.

The ensemble of analyses would be improved by measurements of surface superconductivity and sub-surface contamination. Extraction of a local value for κ within the top few penetration depths would provide refinement of the model.

**Acknowledgments**

This work was supported by the U.S. Department of Energy under contract No. DE-AC02-07CH11359. The authors would like to thank G. Ciovati, J. Clem, A. Gurevich, H. Padamsee, N. Solyak, G. Wu, and V. Yakolev for stimulating discussions. Chemical work was carried out with the kind assistance of D. Burk, D. Hicks, R. Schuessler, and C. Thompson. Profilometry measurements were aided by C. Cooper, C. Thompson, and M. Ge.